\documentclass[12pt]{article}

\usepackage[totalheight = 23cm, totalwidth = 17cm]{geometry}
\usepackage{amssymb,amsmath,amsfonts,amsbsy,graphicx}
\usepackage{bm,simplewick,axodraw}

\newcommand{\mpl}{m_{\rm Pl}}

\newcommand{\calB}{{\cal B}}

\newcommand{\calG}{{\cal G}}

\newcommand{\calP}{{\cal P}}
\newcommand{\calR}{{\cal R}}

\begin{document}

\begin{titlepage}

\begin{center}

\vspace*{-10ex}
\hspace*{\fill}
CERN-PH-TH/2010-267

\vskip 1.5cm

\Huge{Non-linear corrections to \\ inflationary power spectrum}

\vskip 1cm

\large{
Jinn-Ouk Gong$^{*,\dag}$\footnote{jinn-ouk.gong@cern.ch}
\hspace{0.2cm}
Hyerim Noh$^{\ddag}$\footnote{hr@kasi.re.kr}
\hspace{0.2cm}\mbox{and}\hspace{0.2cm}
Jai-chan Hwang$^{\S}$\footnote{jchan@knu.ac.kr}
\\
\vspace{0.5cm}
{\em
${}^*$ Instituut-Lorentz for Theoretical Physics, Universiteit Leiden
\\
2333 CA Leiden, The Netherlands
\\
\vspace{0.2cm}
${}^\dag$ Theory Division, CERN
\\
CH-1211 Gen\`eve 23, Switzerland
\\
\vspace{0.2cm}
${}^\ddag$ Korea Astronomy and Space Science Institute
\\
Daejeon 305-348, Republic of Korea
\\
\vspace{0.2cm}
${}^\S$ Department of Astronomy and Atmospheric Sciences
\\
Kyungpook National University
\\
Daegu 702-701, Republic of Korea}
}

\vskip 0.5cm

\today

\vskip 1.2cm

\end{center}

\begin{abstract}

We study non-linear contributions to the power spectrum of the
curvature perturbation on super-horizon scales, produced during
slow-roll inflation driven by a canonical single scalar field. We
find that on large scales the linear power spectrum
dominates and leading non-linear corrections remain
negligible, indicating that we can safely rely on linear
perturbation theory to study inflationary power spectrum. We also
briefly comment on the infrared and ultraviolet behaviour of the
non-linear corrections.

\end{abstract}

\end{titlepage}

\setcounter{page}{0}
\newpage
\setcounter{page}{1}

\section{Introduction}
\label{sec:intro}

It is now widely accepted that primordial cosmic inflation in the
early universe~\cite{inflation} is the leading candidate
to provide both the initial conditions for successful hot big bang
universe~\cite{Lyth:1998xn}, and a natural mechanism of generating
the primordial perturbations. The quantum fluctuations of the scalar
field which drives inflation, the ``inflaton'' field, are stretched
to super-horizon scales during inflation and become the seeds of
temperature anisotropies of the cosmic microwave background and
large scale inhomogeneities~\cite{inflation_spectrum}. The
computation of the power spectrum of these primordial perturbations
has become a well established subject~\cite{Mukhanov:1990me}.

Although this linear picture is well studied, its extension to
include non-linear effect has not been studied seriously. Only
recently, computations to second order metric and matter
perturbations were carried out~\cite{2ndorder}, reporting divergent
behaviors of non-linear perturbations dominating over linear order
ones. This implies that perturbation theory breaks down in the seed
generation stage.
However, this is not likely to be the case considering the fact that
the observed power spectrum on large scales, which is supposed to
reflect the primeval behavior of quantum fluctuations, is as small
as $10^{-9}$~\cite{observations}.

We should note that to consistently study the effects of non-linear
corrections to the power spectrum, one needs third order
perturbations. This means the second order perturbation of previous
studies~\cite{2ndorder} is {\em not} sufficient. The non-linear
contributions to the power spectrum in terms of other correlation
functions have been known in terms of the $\delta{N}$
formalism~\cite{deltaNnonlinear}, but it is not clear if those
results also exhibit any breakdown of perturbation theory.

In this note, we carry out explicit computations of the leading
non-linear contributions to the power spectrum. We consider the
simplest but important case of single field slow-roll inflation. The
metric and scalar field matter perturbations can be described in
terms of the gauge invariant comoving curvature perturbation ${\cal
R}$. We further take both large scale limit and slow-roll
approximation, which greatly simplify the calculations yet give the
leading order result. Our result shows that, contrary to the
previous report~\cite{2ndorder}, the linear power spectrum dominates
completely and the non-linear contributions are negligible.
This assures the validity of linear cosmological perturbation theory
in handling the quantum generation process during inflation.

This note is outlined as follows.
In Section~\ref{sec:equations} we present the non-linear equation of
the curvature perturbation in the comoving gauge. In
Section~\ref{sec:solutions} we solve the equation order by order and
find the non-linear solutions up to third order. In
Section~\ref{sec:powerspectrum} we calculate the contributions to the
power spectrum from non-linear corrections. In
Section~\ref{sec:conclusions} we present the conclusion.

\section{Equations}
\label{sec:equations}

\subsection{Setup}
\label{subsec:setup}

Our starting point is the action of gravity with a minimally coupled
scalar field,
\begin{equation}\label{action}
S = \int d^4x N\sqrt{\gamma} \left\{ \frac{\mpl^2}{2} \left( R^{(3)}
+ K^i{}_jK^j{}_i - K^2 \right) + \frac{1}{2} \left[ \frac{(\phi_{,0}
- N^i\phi_{|i})^2}{N^2} - \phi^{|i}\phi_{|i} \right] - V(\phi)
\right\} \, ,
\end{equation}
where the action is written using the Arnowitt-Deser-Misner (ADM) metric~\cite{Arnowitt:1962hi},
\begin{equation}\label{ADM_metric}
ds^2 = -N^2 (dx^{0})^2 + \gamma_{ij} (N^i d x^0 + d x^i)(N^j d x^0 +
d x^j) \, .
\end{equation}
Vertical bars denote 3-space covariant derivatives, $R^{(3)}$ is the
3-space curvature constructed from $\gamma_{ij}$, and the extrinsic
curvature 3-tensor is given by
\begin{equation}\label{ext_curvature}
K_{ij} = \frac{1}{2N} \left( N_{i|j} + N_{j|i} - \gamma_{ij,0}
\right) \, .
\end{equation}
Below, we denote the traceless part by an overbar,
\begin{equation}\label{tracelessKij}
\overline{K}_{ij} \equiv K_{ij} - \frac{1}{3}\gamma_{ij}K \, ,
\end{equation}
with $K = K^i{}_i = \gamma^{ij}K_{ij}$.

By varying the action (\ref{action}) with respect to $N$, $N^i$ and $\phi$, we obtain respectively the energy and momentum constraint equations, and the equation of motion of $\phi$ as
\begin{align}
\label{energy_const}
\overline{K}^i{}_j\overline{K}^j{}_i -
\frac{2}{3}K^2 - R^{(3)} + \frac{2}{\mpl^2} E = & 0 \, ,
\\
\label{momentum_const}
\overline{K}^j{}_{i|j} - \frac{2}{3}K_{|i} +
\frac{1}{\mpl^2} J_i= & 0 \, ,
\\
\label{field_eq}
\frac{1}{N} \left(\dot\Pi^\phi - N^i\Pi^\phi{}_{|i}
\right) - K\Pi^\phi - \frac{N_{|i}\phi^{|i}}{N} + \phi^{|i}{}{}_{|i}
+ \frac{\partial V}{\partial \phi} = & 0 \, ,
\end{align}
where the conjugate momentum, energy density and momentum of $\phi$
are respectively given by
\begin{align}
\Pi^\phi = & \frac{\phi_{,0} - N^i\phi_{|i}}{N} \, ,
\\
E = & \frac{1}{2} \left[ \left( \Pi^\phi \right)^2 + \phi^{|i}\phi_{|i} \right] + V(\phi) \, ,
\\
J_i = & \Pi^\phi\phi_{|i} \, .
\end{align}

To the second and higher order perturbations, in general, we have couplings among the scalar, vector and tensor perturbations. In this work, we consider only scalar perturbations in a flat Friedmann
background model. As the temporal gauge (hypersurface) condition, we
take the uniform field gauge such that the perturbed scalar field
vanishes, $\delta \phi = 0$. This is the same as the comoving gauge in
the single component case which sets $J_i = 0$. We take a spatial
gauge condition which fixes the spatial gauge degree of freedom
completely~\cite{Bardeen-1988}: under this gauge condition we have
$\gamma_{ij} = a^2 ( 1 + 2 \calR ) \delta_{ij}$,
where $\calR$ is related to the perturbed part of the spatial
curvature $R^{(3)}$. Under our temporal comoving gauge, we call
${\cal R}$ as the ``comoving curvature perturbation''.
These gauge
conditions fix the gauge degrees of freedom completely even to
non-linear orders~\cite{Noh:2004bc}. To summarize, our gauge
conditions are
\begin{align}
   \delta \phi = & 0 \, ,
   \\
   \gamma_{ij} = & a^2 ( 1 + 2 \calR ) \delta_{ij} \, .
\end{align}
(\ref{energy_const}), (\ref{momentum_const}) and (\ref{field_eq}) together with the trace part of
(\ref{ext_curvature}) provide a complete set of equations to
have a closed form equation for the curvature perturbation
$\calR$. In our gauge, (\ref{field_eq}) and (\ref{energy_const})
become
\begin{align}
   - {\ddot \phi \over N^2}
       + \left( K + {\dot N \over N^2}
       - {N_{|i} \over N^2} N^i \right)
       {\dot \phi \over N} = & \frac{\partial V}{\partial \phi} \, ,
   \label{eq-1}
   \\
   \overline{K}^{ij} \overline{K}_{ij}
       - \frac{2}{3} K^2
       + \left( {\dot \phi^2 \over {2N^2}} + V \right) = & R^{(3)} \,
       ,
   \label{eq-2}
\end{align}
where an overdot denotes a time derivative with $x^0 = t$. By
removing $K$ in these equations we can derive a relation between
$\calR$ in $R^{(3)}$ and $\dot N$. Further, (\ref{ext_curvature})
and (\ref{momentum_const}) become
\begin{align}
   K = & {1 \over 2 N} \left( 2 N^i{}_{|i}
       - \gamma^{ij} \dot \gamma_{ij} \right) \, ,
   \label{eq-3}
   \\
   K_{|i} = & {3 \over 2} \overline{K}^j{}_{i|j} \, .
   \label{eq-4}
\end{align}
By removing $K$ in these equations, we can derive another relation
between $\dot \calR$ in $\dot \gamma_{ij}$ and $N$. Combining the
two relations between $\dot \calR$ and $\dot N$ we can derive a
closed form equation of $\ddot \calR$ even to non-linear orders in
perturbations.

\subsection{Non-linear equations}
\label{subsec:nonlinear_equations}

Now we consider non-linear perturbation theory of
$\calR$~\cite{Noh:2004bc}. We can explicitly combine (\ref{eq-1}),
(\ref{eq-2}), (\ref{eq-3}) and (\ref{eq-4}) to write the equation
purely using the curvature perturbation $\calR$ to all perturbation
orders. We will write up to third order, since we are interested in
the next-to-leading corrections to the power spectrum.
Although the full non-linear equation of $\cal R$
is very lengthy,
we can gain more control by taking two approximations. We are
interested in the behaviour of $\calR$ on large scales.
Thus we keep the leading correction terms in the large scale limit,
which include two spatial derivatives. This corresponds to taking the
super-horizon limit. Further, in the single field inflation model we consider,
the slow-roll approximation is valid with a tiny deviation of
$\mathcal{O}(\epsilon)$, where $\epsilon = -\dot{H}/H^2 =
(\dot\phi/H)^2/(2\mpl^2)$ is the slow-roll parameter. The zeroth
order terms in $\epsilon$ give the leading contributions and we
keep only those terms.

From (\ref{eq-1}), (\ref{eq-2}), (\ref{eq-3}) and (\ref{eq-4}), up to third
order we have
\begin{align}\label{3th_order_eq_ls}
\ddot\calR + 3H\dot\calR - \frac{\Delta}{a^2}\calR = & \frac{1}{a^2} \bigg[ -\frac{7}{4}\calR^{,i}\calR_{,i} - 2\calR\Delta\calR - \frac{1}{2}\Delta^{-1} \left( \calR^{,i}\Delta\calR \right)_{,i}
\nonumber\\
& \hspace{0.5cm} + \frac{13}{2}\calR\calR^{,i}\calR_{,i} + 4\calR^2\Delta\calR -
\calR \Delta^{-1} \left( \calR^{,i}\Delta\calR \right)_{,i}
\nonumber\\
& \hspace{0.5cm} + \frac{1}{2}\Delta^{-1} \left( 4\calR\calR^{,i}\Delta\calR +
2\calR^{,j}\calR_{,j}\calR^{,i} + \Delta\calR\Delta^{-1}\calB_2^{,i}
+ \calR^{,ij}\Delta^{-1}\calB_{2,j} \right)_{,i} \bigg] \, ,
\end{align}
where $\Delta \equiv \delta^{ij}\partial_i\partial_j$ and $\Delta^{-1}$ are the Laplacian and inverse Laplacian operators, and
\begin{equation}
\calB_2 = \frac{7}{4}\calR^{,i}\calR_{,i} + 2\calR\Delta\calR - \frac{3}{2}\Delta^{-1} \left( \calR^{,i}\Delta\calR \right)_{,i} \, .
\end{equation}

We can write (\ref{3th_order_eq_ls}) in the Fourier space by introducing the Fourier component of $\calR$ as
\begin{equation}
\calR = \int \frac{d^3k}{(2\pi)^3} e^{i{\bm k}\cdot{\bm{x}}} \calR_{\bm k} \, .
\end{equation}
Then, we can find
\begin{align}\label{Fourier_eq}
\ddot\calR_{\bm k} + 3H\dot\calR_{\bm k} + \frac{k^2}{a^2}\calR_{\bm k} = & \frac{1}{a^2} \int \frac{d^3q_1d^3q_2}{(2\pi)^3} \delta^{(3)}({\bm k} - {\bm q}_{12}) \calR_{{\bm q}_1}\calR_{{\bm q}_2} \left( 2q_2^2 + \frac{7}{4}{\bm q}_1\cdot{\bm q}_2 + \frac{q_2^2}{2k^2}{\bm k}\cdot{\bm q}_1 \right)
\nonumber\\
& + \frac{1}{a^2} \int \frac{d^3q_1d^3q_2d^3q_3}{(2\pi)^{3\cdot2}} \delta^{(3)}({\bm k}-{\bm q}_{123}) \calR_{{\bm q}_1}\calR_{{\bm q}_2}\calR_{{\bm q}_3}
\nonumber\\
& \hspace{1cm} \times \left[ -4q_3^2 - \frac{13}{2}{\bm q}_2\cdot{\bm q}_3 -
\frac{2q_3^2}{k^2}{\bm k}\cdot{\bm q}_2 - \frac{{\bm k}\cdot{\bm
q_3}}{k^2}{\bm q}_1\cdot{\bm q}_2 + \frac{q_3^2}{q_{23}^2}{\bm
q}_{23}\cdot{\bm q}_2 \right.
\nonumber\\
& \hspace{1.5cm} \left. + \frac{1}{2} \left( \frac{q_1^2}{q_{23}^2}\frac{{\bm
k}\cdot{\bm q}_{23}}{k^2} + \frac{{\bm k}\cdot{\bm q}_1}{k^2}
\frac{{\bm q}_1\cdot{\bm q}_{23}}{q_{23}^2} \right) \left(
-\frac{7}{4}{\bm q}_2\cdot{\bm q}_3 - 2q_3^2 +
\frac{3q_3^2}{2q_{23}^2}{\bm q}_{23}\cdot{\bm q}_2 \right) \right]
\, ,
\end{align}
where we have introduced a shorthanded notation ${\bm q}_{12\cdots n} = {\bm q}_1 + {\bm q}_2 + \cdots + {\bm q}_n$. The non-linear parts of (\ref{3th_order_eq_ls}) and (\ref{Fourier_eq}) are valid to the leading order in the large scale approximation.
We note that the second and third order terms in the right hand sides of (\ref{3th_order_eq_ls}) and (\ref{Fourier_eq}) have $\Delta/a^2$ order factor, thus suppressed in the large scale limit.

\section{Solutions}
\label{sec:solutions}

We can find the solution of (\ref{Fourier_eq}) by perturbative
expansion. We first consider the linear solution $\calR_{\bm k}^{(1)}$
of (\ref{Fourier_eq}), which satisfies
\begin{equation}
\ddot\calR_{\bm k}^{(1)} + 3H\dot\calR_{\bm k}^{(1)} + \frac{k^2}{a^2}\calR_{\bm k}^{(1)} = 0 \, .
\end{equation}
On large scales, we find
\begin{equation}\label{linear_sol}
\calR_{\bm k}^{(1)} = C_1 + \frac{C_2}{a^3}  \, ,
\end{equation}
with $C_1$ and $C_2$ being integration constants depending only on
${\bm k}$. In an expanding phase we are interested in, the constant
$C_1$ is (relatively) the growing solution, while $C_2/a^3$ is the
decaying one. Thus, we neglect the transient decaying solution when
we feedback the linear solution to obtain non-linear contributions.
The coefficient $C_1$ will be determined from quantum fluctuations
as we will see in the next section.

With the linear solution $\calR_{\bm k}^{(1)} = C_1 ({\bm k})$, we
perturbatively expand the full non-linear solution $\calR_{\bm k}$
in terms of momentum dependent symmetric kernels as\footnote{See
also Ref.~\cite{Jeong:2010ag} for the perturbative solution of the
density contrast $\delta$ in Einstein-de Sitter universe.}
\begin{align}\label{Rexpansion}
\calR_{\bm k} = & \sum_{n=1}^\infty \int \frac{d^3q_1 \cdots d^3q_n}{(2\pi)^{3(n-1)}} \delta^{(3)}({\bm k} - {\bm q}_{1\cdots n}) \calG_n^{(s)}({\bm q}_1, \cdots {\bm q}_n) \calR_{{\bm q}_1}^{(1)} \cdots \calR_{{\bm q}_n}^{(1)}
\nonumber\\
= & \calR_{\bm k}^{(1)} + \calR_{\bm k}^{(2)} + \calR_{\bm k}^{(3)} + \cdots \, ,
\end{align}
where $\calG_1(\bm k) = 1$. Using (\ref{Rexpansion}), the original non-linear equation
(\ref{Fourier_eq}) is reduced to simple differential equations of
the kernels $\calG_n$ order by order.

Plugging (\ref{Rexpansion}) into (\ref{Fourier_eq}), at second order we have
\begin{equation}
\ddot\calG_2 + 3H\dot\calG_2 = \frac{1}{a^2} \left( 2q_2^2 + \frac{7}{4}{\bm q}_1\cdot{\bm q}_2 + \frac{q_2^2}{2k^2}{\bm k}\cdot{\bm q}_1 \right) \, .
\end{equation}
This can be solved to give
\begin{equation}\label{sol-G2}
\calG_2({\bm q}_1,{\bm q}_2) = -\frac{1}{2a^2H^2} \left( 2q_2^2 + \frac{7}{4}{\bm q}_1\cdot{\bm q}_2 + \frac{q_2^2}{2k^2}{\bm k}\cdot{\bm q}_1 \right) \, .
\end{equation}
The symmetrized kernel $\calG_2^{(s)}({\bm q}_1,{\bm q}_2)$ which we
will use to find higher order kernels is then given by exchanging
the arguments,
\begin{equation}\label{G2s}
\calG_2^{(s)}({\bm q}_1,{\bm q}_2) = \frac{1}{2!} \left[ \calG_2({\bm q}_1,{\bm q}_2) + \calG_2({\bm q}_2,{\bm q}_1) \right] \, .
\end{equation}
With the second order kernel, from (\ref{Rexpansion}) we can write
the second order solution $\calR_{\bm k}^{(2)}$ as
\begin{equation}\label{R2}
\calR_{\bm k}^{(2)} = \int \frac{d^3q_1d^3q_2}{(2\pi)^3} \delta^{(3)}({\bm k}-{\bm q}_{12}) \calG_2^{(s)}({\bm q}_1,{\bm q}_2) \calR_{{\bm q}_1}^{(1)} \calR_{{\bm q}_2}^{(1)} \, .
\end{equation}

For the third order kernel $\calG_3$ we have
\begin{align}
\ddot\calG_3 + 3H\dot\calG_3 = & \frac{1}{a^2} \left\{ \calG_2^{(s)}({\bm q}_2,{\bm q}_3) \left[ 2 \left( q_1^2 + q_{23}^2 \right) + \frac{7}{2}{\bm q}_1\cdot{\bm q}_{23} + \frac{q_1^2}{2k^2}{\bm k}\cdot{\bm q}_{23} + \frac{q_{23}^2}{2k^2}{\bm k}\cdot{\bm q}_1 \right] \right.
\nonumber\\
& \hspace{0.5cm} + \left[ -4q_3^2 - \frac{13}{2}{\bm q}_2\cdot{\bm q}_3 - \frac{2q_3^2}{k^2}{\bm k}\cdot{\bm q}_2 - \frac{{\bm k}\cdot{\bm q_3}}{k^2}{\bm q}_1\cdot{\bm q}_2 + \frac{q_3^2}{q_{23}^2}{\bm q}_{23}\cdot{\bm q}_2 \right.
\nonumber\\
& \hspace{1cm} \left.\left. + \frac{1}{2} \left( \frac{q_1^2}{q_{23}^2}\frac{{\bm k}\cdot{\bm q}_{23}}{k^2} + \frac{{\bm k}\cdot{\bm q}_1}{k^2} \frac{{\bm q}_1\cdot{\bm q}_{23}}{q_{23}^2} \right) \left( -\frac{7}{4}{\bm q}_2\cdot{\bm q}_3 - 2q_3^2 + \frac{3q_3^2}{2q_{23}^2}{\bm q}_{23}\cdot{\bm q}_2 \right) \right] \right\} \, .
\end{align}
Here, let us split the third order kernel $\calG_3$ into two parts,
\begin{equation}
\calG_3({\bm q}_1,{\bm q}_2,{\bm q}_3) = \calG_{31}({\bm q}_1,{\bm q}_2,{\bm q}_3) + \calG_{32}({\bm q}_1,{\bm q}_2,{\bm q}_3) \, ,
\end{equation}
where $\calG_{31}$ denotes the terms multiplied by the second order
kernel $\calG_2^{(s)}$ and $\calG_{32}$ the rest. For $\calG_{31}$,
we can find that $\calG_{31}$ is a product of two $\calG_2^{(s)}$'s as
\begin{equation}
\calG_{31}({\bm q}_1,{\bm q}_2,{\bm q}_3) = -\calG_2^{(s)}({\bm q}_2,{\bm q}_3) \calG_2^{(s)}({\bm q}_1,{\bm q}_{23}) \, .
\end{equation}
We can also find $\calG_{32}$ as
\begin{align}
\calG_{32}({\bm q}_1,{\bm q}_2,{\bm q}_3) = & -\frac{1}{2a^2H^2} \left[ -4q_3^2 - \frac{13}{2}{\bm q}_2\cdot{\bm q}_3 - \frac{2q_3^2}{k^2}{\bm k}\cdot{\bm q}_2 - \frac{{\bm k}\cdot{\bm q_3}}{k^2}{\bm q}_1\cdot{\bm q}_2 + \frac{q_3^2}{q_{23}^2}{\bm q}_{23}\cdot{\bm q}_2 \right.
\nonumber\\
& \hspace{1.8cm} \left. + \frac{1}{2} \left( \frac{q_1^2}{q_{23}^2}\frac{{\bm
k}\cdot{\bm q}_{23}}{k^2} + \frac{{\bm k}\cdot{\bm q}_1}{k^2}
\frac{{\bm q}_1\cdot{\bm q}_{23}}{q_{23}^2} \right) \left(
-\frac{7}{4}{\bm q}_2\cdot{\bm q}_3 - 2q_3^2 +
\frac{3q_3^2}{2q_{23}^2}{\bm q}_{23}\cdot{\bm q}_2 \right) \right]
\, .
\end{align}
Then the symmetrized third order kernel is written as
\begin{equation}
\calG_3^{(s)}({\bm q}_1,{\bm q}_2,{\bm q}_3) = \frac{1}{3!} \left[ \calG_3({\bm q}_1,{\bm q}_2,{\bm q}_3) + \mbox{5 permutations} \right] \, ,
\end{equation}
and the third order solution is given by
\begin{equation}\label{R3}
\calR_{\bm k}^{(3)} = \int \frac{d^3q_1d^3q_2d^3q_3}{(2\pi)^{6}}
\delta^{(3)}({\bm k}-{\bm q}_{123}) \calG_3^{(s)}({\bm q}_1,{\bm
q}_2,{\bm q}_3) \calR_{{\bm q}_1}^{(1)} \calR_{{\bm q}_2}^{(1)}
\calR_{{\bm q}_3}^{(1)} \, .
\end{equation}

\section{Power spectrum}
\label{sec:powerspectrum}

\begin{figure}[h]
\begin{center}
 \begin{picture}(380,50)(0,0)
  \Line(0,30)(15,30)
  \Vertex(15,30){2}
   \Line(15,30)(45,30)
  \Vertex(45,30){2}
  \Line(45,30)(60,30)
   \Text(30,10)[]{$\mathcal{P}_\mathcal{R}(k)$}
   \Text(70,30)[]{$=$}
  \Line(80,30)(95,30)
  \Vertex(95,30){2}
   \DashLine(95,30)(125,30){1.8}
  \Vertex(125,30){2}
  \Line(125,30)(140,30)
   \Text(110,10)[]{$\mathcal{P}_\mathcal{R}^{(11)}(k)$}
   \Text(150,30)[]{$+$}
  \Line(160,30)(175,30)
  \Vertex(175,30){2}
   \DashLine(175,30)(190,30){1.8}
   \DashCArc(197.5,30)(7.5,0,360){1.8}
  \Vertex(205,30){2}
  \Line(205,30)(220,30)
   \Text(190,10)[]{$\mathcal{P}_\mathcal{R}^{(12)}(k)$}
   \Text(230,30)[]{$+$}
  \Line(240,30)(255,30)
  \Vertex(255,30){2}
   \DashCArc(270,16)(20,40,140){1.8}
   \DashCArc(270,44)(20,220,320){1.8}
  \Vertex(285,30){2}
  \Line(285,30)(300,30)
   \Text(270,10)[]{$\mathcal{P}_\mathcal{R}^{(22)}(k)$}
   \Text(310,30)[]{$+$}
  \Line(320,30)(335,30)
  \Vertex(335,30){2}
   \DashLine(335,30)(365,30){1.8}
   \DashCArc(335,37.5)(7.5,0,360){1.8}
  \Vertex(365,30){2}
  \Line(365,30)(380,30)
   \Text(350,10)[]{$\mathcal{P}_\mathcal{R}^{(13)}(k)$}
 \end{picture}
\end{center}
\vspace{-1em}
 \caption{Diagramatic representation of the power spectrum up to next-to-leading order corrections. Up to this order, all the corrections include one internal momentum loop integral so that they can be dubbed ``one-loop'' corrections. However, note that $\calP_\calR^{(12)}$, which includes bispectrum, is of higher order than the other two corrections.}
 \label{fig:diagram}
\end{figure}
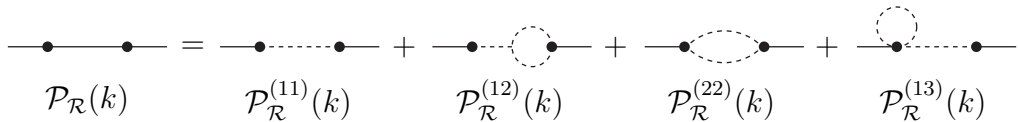

Having found the Fourier mode solution $\calR_\mathbf{k}$, we can
write the power spectrum as
\begin{align}\label{spectrum}
\left\langle \calR_{{\bm k}_1}\calR_{{\bm k}_2} \right\rangle = &
\left\langle \calR_{{\bm k}_1}^{(1)}\calR_{{\bm k}_2}^{(1)} + \left[
\calR_{{\bm k}_1}^{(1)}\calR_{{\bm k}_2}^{(2)} + \calR_{{\bm
k}_1}^{(2)}\calR_{{\bm k}_2}^{(1)} \right] + \left[ \calR_{{\bm
k}_1}^{(2)}\calR_{{\bm k}_2}^{(2)} + \calR_{{\bm
k}_1}^{(1)}\calR_{{\bm k}_2}^{(3)} + \calR_{{\bm
k}_1}^{(3)}\calR_{{\bm k}_2}^{(1)} \right] + \cdots \right\rangle
\nonumber\\
\equiv & (2\pi)^3 \delta^{(3)}({\bm k}_1+{\bm k}_2)
\frac{2\pi^2}{k_1^3}\calP_\calR(k_1)
\nonumber \\
= & (2\pi)^3 \delta^{(3)}({\bm k}_1+{\bm k}_2) \frac{2\pi^2}{k_1^3}
\left\{ \calP^{(11)}_\calR(k_1) + \calP^{(12)}_\calR(k_1) + \left[
\calP^{(22)}_\calR(k_1) + \calP^{(13)}_\calR(k_1) \right] + \cdots
\right\} \, .
\end{align}
Here we have grouped the terms of the same perturbation order.
(\ref{spectrum}) is diagramatically shown in Figure~\ref{fig:diagram}.

Before we proceed, we make some remarks. As mentioned in Section~\ref{sec:intro}, the one-loop corrections to the power spectrum are known from the $\delta{N}$ formalism~\cite{deltaNnonlinear}. Especially, the structure of these corrections in terms of correlation functions which we show below is precisely the same, as it should be. However, there are two differences. First, we obtain our results by solving the non-linear Einstein equation. This provides an alternative approach to non-linearity in the power spectrum. More importantly, we present the explicit momentum dependences of these correction terms, which are hidden in the derivatives of the number of $e$-folds $N$ in the $\delta{N}$ formalism.

\subsection{Linear power spectrum}

In the context of inflationary cosmology, the brackets of
(\ref{spectrum}) are taken with respect to the vacuum state of the
perturbation {\em operator}, in our case $\calR$. Thus, to estimate
the linear power spectrum of $\calR$, we can expand $\calR_{\bm k}$
in terms of the creation and annihilation operators of quantum
harmonic oscillators, namely,
\begin{equation}\label{operator_expansion}
\calR_{\bm k}^{(1)} = \frac{1}{z} \left( a_{\bm k}u_k + a_{-{\bm k}}^\dag u_k^* \right) \, ,
\end{equation}
where $z = a\dot\phi/H$~\cite{Mukhanov-1988} and the creation and
annihilation operators satisfy the canonical commutation relations
\begin{equation}
\left[ a_{\bm k}, a_{\bm q}^\dag \right] = (2\pi)^3\delta^{(3)}({\bm k}-{\bm q}) \, ,
\end{equation}
otherwise zero. Then, the mode function equation of $u_k$ is given by
\begin{equation}\label{uk_eq}
u_k'' + \left( k^2 - \frac{z''}{z} \right)u_k = 0 \, ,
\end{equation}
where a prime denotes a derivative with respect to the conformal
time $d\eta = dt/a$. The solution which satisfies the Bunch-Davies
vacuum boundary condition in the zeroth order slow-roll
approximation is known to be~\cite{Bunch-Davies}
\begin{equation}\label{uk_sol}
u_k = \frac{1}{\sqrt{2k}} \left( 1 + i\frac{aH}{k}
\right)e^{ik/(aH)}
\underset{k\to0}{\longrightarrow}
\frac{iaH}{\sqrt{2k^3}}e^{ik/(aH)} \, .
\end{equation}
As $k\to0$ we have the asymptotic solution $|u_k| \propto z$, which
can be read from (\ref{uk_eq}). This is the constant amplitude of
the growing solution $C_1({\bm k})$ we found in (\ref{linear_sol}).
Note that using the exact solution (\ref{uk_sol}), we can recover the same second order kernel (\ref{sol-G2}) in the large scale limit and slow-roll approximation.

Using (\ref{operator_expansion}) and (\ref{uk_sol}) and
comparing with (\ref{spectrum}), we have
\begin{equation}
\left\langle \calR_{{\bm k}_1}^{(1)}\calR_{{\bm k}_2}^{(1)} \right\rangle = (2\pi)^3\delta^{(3)}({{\bm k}_1}+{{\bm k}_2}) \frac{|u_k|^2}{z^2} = (2\pi)^3\delta^{(3)}({{\bm k}_1}+{{\bm k}_2}) \frac{2\pi^2}{k_1^3}\calP_\calR^{(11)}(k_1) \, ,
\end{equation}
so that we can find the linear power spectrum as
\begin{equation}\label{P11}
\calP_\calR^{(11)}(k) = \frac{k^3}{2\pi^2}\frac{|u_k|^2}{z^2} = \left( \frac{H}{2\pi} \right)^2 \left( \frac{H}{\dot\phi} \right)^2 \, ,
\end{equation}
which is the well-known scale invariant power spectrum produced
during inflation~\cite{inflation_spectrum}. In our approximation
there is no momentum dependence for $\calP_\calR^{(11)}$.

\subsection{Leading corrections to the power spectrum}

Next, we consider the leading correction to the linear power spectrum, the terms inside the first square brackets of (\ref{spectrum}). From (\ref{R2}), we have
\begin{align}\label{P12}
\left\langle \calR_{{\bm k}_1}^{(1)}\calR_{{\bm k}_2}^{(2)} \right\rangle = & \int \frac{d^3q_1d^3q_2}{(2\pi)^3} \delta^{(3)}({\bm k}_2-{\bm q}_{12}) \calG_2^{(s)}({\bm q}_1,{\bm q}_2) \left\langle \calR_{{\bm k}_1}^{(1)}\calR_{{\bm q}_1}^{(1)}\calR_{{\bm q}_2}^{(1)} \right\rangle
\nonumber\\
= & \int \frac{d^3q_1d^3q_2}{(2\pi)^3} \delta^{(3)}({\bm k}_2-{\bm q}_{12}) \calG_2^{(s)}({\bm q}_1,{\bm q}_2) \frac{1}{z^3}
\nonumber\\
& \times \left\langle \left( a_{{\bm k}_1}u_{k_1} + a_{-{\bm k}_1}^\dag u_{k_1}^* \right) \left( a_{{\bm q}_1}u_{q_1} + a_{-{{\bm q}_1}}^\dag u_{q_1}^* \right) \left( a_{{\bm q}_2}u_{q_2} + a_{-{\bm q}_2}^\dag u_{q_2}^* \right) \right\rangle \, .
\end{align}
What we can see immediately is that, we have the combinations of three creation and/or annihilation operators, such as $a_{{\bm k}_1}a_{{\bm q}_1}a_{{\bm q}_2}$. Thus, we have one remaining creation or annihilation operator after using the commutator relations for two of them, which vanishes since it is sandwiched between the vacuum states. Thus,
\begin{equation}
\left\langle \calR_{{\bm k}_1}^{(1)}\calR_{{\bm k}_2}^{(2)} \right\rangle = \left\langle \calR_{{\bm k}_1}^{(2)}\calR_{{\bm k}_2}^{(1)} \right\rangle = 0 \, .
\end{equation}
Hence,
\begin{equation}
\calP_\calR^{(12)} = 0 \, ,
\end{equation}
and non-vanishing corrections appear only in the next order.

Before we move to the next-to-leading order corrections, we consider $\calP_\calR^{(12)}$ further. What we can first note from (\ref{P12}) is that it is sourced by the primordial bispectrum,
\begin{equation}
\left\langle \calR_{{\bm k}_1}^{(1)} \calR_{{\bm k}_2}^{(1)} \calR_{{\bm k}_3}^{(1)} \right\rangle \equiv (2\pi)^3\delta^{(3)}({\bm k}_1+{\bm k}_2+{\bm k}_3) B_\calR({\bm k}_1,{\bm k}_2,{\bm k}_3) \, .
\end{equation}
If the distribution of $\calR_{\bm k}^{(1)}$ is not Gaussian, in general we have a non-vanishing bispectrum. Indeed even for single field slow-roll inflation this is the case~\cite{Maldacena:2002vr}. This intrinsic non-Gaussianity comes from the contributions around the moment of horizon crossing, which we do not take into account by construction~\cite{horizoncrossing}. Moreover, although the leading and the next-to-leading order corrections all include one internal momentum integral so that they are usually classified as one-loop corrections, clearly $\calP_\calR^{(12)}$ is of higher order.

\subsection{Next-to-leading corrections to the power spectrum}

Now we consider the next-to-leading order corrections to the power
spectrum. As we can see from (\ref{spectrum}), there are two
contributions. One is given by the quadratic combination of the
second order solution, and the other by the product of linear and
third order solutions.

\subsubsection{$\calP_\calR^{(22)}$}

Using the second order solution (\ref{R2}), we can find
\begin{align}
\left\langle \calR_{{\bm k}_1}^{(2)}\calR_{{\bm k}_2}^{(2)} \right\rangle = & \int \frac{d^3q_1d^3q_2}{(2\pi)^3} \delta^{(3)}({\bm k}_1-{\bm q}_{12}) \calG_2^{(s)}({\bm q}_1,{\bm q}_2) \int \frac{d^3q_3d^3q_4}{(2\pi)^3} \delta^{(3)}({\bm k}_2-{\bm q}_{34}) \calG_2^{(s)}({\bm q}_3,{\bm q}_4)
\nonumber\\
& \times \left\langle \calR_{{\bm q}_1}^{(1)}\calR_{{\bm q}_2}^{(1)} \calR_{{\bm q}_3}^{(1)}\calR_{{\bm q}_4}^{(1)} \right\rangle \, .
\end{align}
The physically relevant correlations can be understood in terms of
contractions. Since we are correlating {\em different}
perturbations, we have two different ways of contraction as
\begin{equation}
\left\langle \calR_{{\bm k}_1}^{(2)}\calR_{{\bm k}_2}^{(2)} \right\rangle =
 \Big\langle
 \contraction
  {\calR} {{}_{{\bm q}_1}^{(1)}} {\calR_{{\bm q}_2}^{(1)})(\calR} {{}_{{\bm q}_3}^{(1)}}
 \contraction[2ex]
  {{}^{(1)}\calR_{{\bm q}}} {{}_{{}_2}^{(1)}} {)(\calR_{{\bm q}_3}^{(1)}\calR_{{\bm q}}} {{}_{{}_4}^{(1)}}
  \Big( \calR_{{\bm q}_1}^{(1)}\calR_{{\bm q}_2}^{(1)} \Big)
  \Big( \calR_{{\bm q}_3}^{(1)}\calR_{{\bm q}_4}^{(1)} \Big)
 \Big\rangle +
 \Big\langle
 \contraction
  {\calR} {{}_{{\bm q}_1}^{(1)}} {\calR_{{\bm q}_2}^{(1)})(\calR_{{\bm q}_3}^{(1)}\calR} {{}_{{\bm q}_4}}
 \contraction[2ex]
  {{}^{(1)}\calR_{\bm q}} {{}_{{}_2}^{(1)}} {)(} {{\calR}_{{\bm q}_3}^{(1)}\calR_{{\bm q}_4}}
  \Big( \calR_{{\bm q}_1}^{(1)}\calR_{{\bm q}_2}^{(1)} \Big)
  \Big( \calR_{{\bm q}_3}^{(1)}\calR_{{\bm q}_4}^{(1)} \Big)
 \Big\rangle \, .
\end{equation}
They correspond to the {\em connected} diagram, the third one in
Figure~\ref{fig:diagram}. Meanwhile, the remaining contractions are
within the {\em same} perturbations and thus irrelevant: we are
interested in the correlation between different perturbations.
This corresponds to a {\em disconnected} diagram,
\begin{equation}
\Big\langle
 \contraction
  {\calR} {{}_{{\bm q}_1}^{(1)}} {} {\calR_{{\bm q}_2}^{(1)}}
 \contraction
  {(\calR_{{\bm q}_1}^{(1)}\calR_{{\bm q}_2}^{(1)})(\calR} {{}_{{\bm q}_3}^{(1)}} {{}} {\calR_{{\bm q}_4}^{(1)})}
  \Big( \calR_{{\bm q}_1}^{(1)}\calR_{{\bm q}_2}^{(1)} \Big)
  \Big( \calR_{{\bm q}_3}^{(1)}\calR_{{\bm q}_4}^{(1)} \Big)
\Big\rangle \mapsto
\parbox[c]{2.9cm}{
 \begin{picture}(80,16)
  \Line(0,8)(20,8)
  \Vertex(20,8){2}
   \DashCArc(28,8)(8,0,360){1.8}
   \DashCArc(52,8)(8,0,360){1.8}
  \Vertex(60,8){2}
  \Line(60,8)(80,8)
 \end{picture}
 } \, .
\end{equation}

Using the above contractions, we obtain
\begin{equation}\label{P22}
\calP_\calR^{(22)}(k) = \frac{k^3}{2\pi} \int d^3q \calG_2^{(s)}({\bm q},{\bm k}-{\bm q}) \calG_2^{(s)}(-{\bm q},-{\bm k}+{\bm q}) \frac{\calP_\calR^{(11)}(q)}{q^3} \frac{\calP_\calR^{(11)}(|{\bm k}-{\bm q}|)}{|{\bm k}-{\bm q}|^3} \, .
\end{equation}
As $\calP_\calR^{(11)}$ is a constant in our approximation, the two
linear power spectra inside can be pulled out of the integral. We
introduce the magnitude of ${\bm q}$ and the cosine $\mu$ between
$\bm q$ and $\bm k$ as $q = rk \; (0\leq r \leq\infty)$ and ${\bm
k}\cdot{\bm q} = k^2r\mu \; (-1\leq \mu \leq1)$.
After the angular integrations, we have
\begin{align}\label{P22rint}
\calP_\calR^{(22)}(k) = & \frac{1}{420} \left( \frac{k}{aH} \right)^4 \left[ \calP_\calR^{(11)} \right]^2 \int_0^\infty \frac{dr}{r^2|1-r||1+r|}
\nonumber\\
& \times \Big[ \left( -41 + 64r + 14r^2 + 84r^3 + 70r^4 + 14r^5 +
14r^6 - 6r^7 - 6r^8 \right) |1-r|
\nonumber\\
& \hspace{0.5cm} + \left( 41 + 64r - 14r^2 + 84r^3 - 70r^4 + 14r^5 - 14r^6 - 6r^7 +
6r^8 \right) |1+r| \, \Big] \, .
\end{align}
We have overall momentum dependence as $\calP_\calR^{(22)} \propto k^4$.

\subsubsection{$\calP_\calR^{(13)}$}

Next, we move to the rest two terms. We find that
\begin{align}
\left\langle \calR_{{\bm k}_1}^{(1)}\calR_{{\bm k}_2}^{(3)} \right\rangle = & \int \frac{d^3q_1d^3q_2d^3q_3}{(2\pi)^6} \delta^{(3)}({\bm k}_2-{\bm q}_{123}) \calG_3^{(s)}({\bm q}_1,{\bm q}_2,{\bm q}_3) \left\langle \calR_{{\bm k}_1}^{(1)} \calR_{{\bm q}_1}^{(1)} \calR_{{\bm q}_2}^{(1)} \calR_{{\bm q}_3}^{(1)} \right\rangle \, .
\end{align}
We can collect relevant correlations as before using contraction.
Here, we are correlating a single $\calR^{(1)}$ to one
$\calR^{(3)}$, which contains three $\calR^{(1)}$'s. Thus
all the possible combinations of contractions include cross
correlations. That is,
\begin{align}
\left\langle \calR_{{\bm k}_1}^{(1)} \left( \calR_{{\bm q}_1}^{(1)} \calR_{{\bm q}_2}^{(1)} \calR_{{\bm k}_3}^{(1)} \right) \right\rangle = &
 \Big\langle
 \contraction[1.5ex]
  {} {\calR} {{}_{{\bm k}_1}^{(1)}(} {\calR}
 \contraction[1.5ex]
  {\calR_{{\bm k}_1}^{(1)}(\calR_{{\bm q}_1}^{(1)}} {\calR} {{}_{{\bm q}_2}^{(1)}} {\calR}
 \calR_{{\bm k}_1}^{(1)} \Big( \calR_{{\bm q}_1}^{(1)} \calR_{{\bm q}_2}^{(1)} \calR_{{\bm q}_3}^{(1)} \Big)
 \Big\rangle +
 \Big\langle
 \contraction[1.5ex]
  {} {\calR} {{}_{{\bm k}_1}^{(1)}(\calR_{{\bm q}_1}^{(1)}} {\calR_{{\bm q}}}
 \contraction[2.5ex]
  {\calR_{{\bm k}_1}^{(1)} (} {\calR} {{}_{{\bm q}_1}^{(1)} \calR_{{\bm q}_2}^{(1)}} {\calR}
 \calR_{{\bm k}_1}^{(1)} \Big( \calR_{{\bm q}_1}^{(1)} \calR_{{\bm q}_2}^{(1)} \calR_{{\bm q}_3}^{(1)} \Big)
 \Big\rangle
\nonumber\\
& + \Big\langle
 \contraction[1.5ex]
  {} {\calR} {{}_{{\bm k}_1}^{(1)}(\calR_{{\bm q}_1}^{(1)}\calR_{{\bm q}_2}^{(1)}} {\calR}
 \contraction[2.5ex]
  {\calR_{{\bm k}_1}^{(1)} (} {\calR} {{}_{{\bm q}_1}^{(1)}} {\calR}
 \calR_{{\bm k}_1}^{(1)} \Big( \calR_{{\bm q}_1}^{(1)} \calR_{{\bm q}_2}^{(1)} \calR_{{\bm q}_3}^{(1)} \Big)
 \Big\rangle \, .
\end{align}
Thus, any correlation automatically includes the meaningful one,
i.e. the contractions between {\em different} perturbations. Then,
after some computations, we can find
\begin{equation}
\calP_\calR^{(13)}(k) = \frac{3}{4\pi}k^3 \int d^3q \left[ \calG_3^{(s)}({\bm k},{\bm q},-{\bm q}) + \calG_3^{(s)}(-{\bm k},{\bm q},-{\bm q}) \right] \frac{\calP_\calR^{(11)}(k)}{k^3} \frac{\calP_\calR^{(11)}(q)}{q^3} \, .
\end{equation}
This can be also analytically integrated with respect to angles and we find
\begin{align}\label{P13rint}
\calP_\calR^{(13)}(k) = & \frac{1}{128} \left( \frac{k}{aH}
\right)^4 \left[ \calP_\calR^{(11)} \right]^2 \int_0^\infty
\frac{dr}{r^2} \bigg[ 2r \left( -71 - 97r^2 - 25r^4 + r^6 \right)
\nonumber\\
& \hspace{5.2cm} + \left( 1-r^2 \right)^2 \left( -7 - 2r^2 + r^4 \right) \log \left| \frac{1-r}{1+r} \right| \bigg]
\nonumber\\
& - \frac{1}{16} \left( \frac{k}{aH} \right)^2 \left[
\calP_\calR^{(11)} \right]^2 \int_0^\infty \frac{dr}{r^2} \bigg[ 2r
\left( -30 - 23r^2 + 3r^4 \right) + \left( 2 - 5r^2 + 3r^6 \right)
\log \left| \frac{1-r}{1+r} \right| \bigg]
\nonumber \\
\equiv & \ \calP_\calR^{(13a)}(k) + \calP_\calR^{(13b)}(k)\, .
\end{align}
We have different $k$ dependence for the two terms
$\calP_\calR^{(13a)} \propto k^4$ and $\calP_\calR^{(13b)} \propto
k^2$.

\subsection{Numerical integration}

\begin{figure}[h]
 \begin{center}
  \includegraphics[width=12cm]{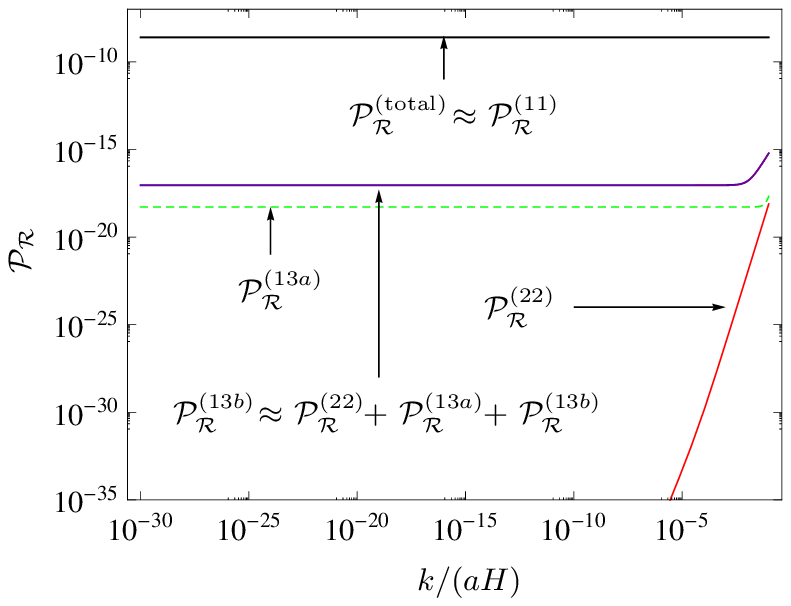}
 \end{center}
 \vspace{-2em}
 \caption{
          Total second order power spectrum and the contributions from non-linear
          corrections are shown for scales between the horizon scale to $10^{30}$ factor larger than the horizon scale at, say, the end of inflation; this covers the scales relevant to current observations.
          Note that $\calP_\calR^{(13a)}$ has negative value.
          We take the absolute value of $\calP_\calR^{(13a)}$ which is denoted by a dashed line.
          $\calP_\calR^{(22)}$ continues the slope till the large-scale limit.
          $\calP_\calR^{(22)} + \calP_\calR^{(13a)} + \calP_\calR^{(13b)}$
          nearly coincides with $\calP_\calR^{(13b)}$,
          and $\calP_\calR^{(11)}$ nearly coincides with $\calP_\calR^{({\rm total})}$.
          }
 \label{fig:Ptotal}
\end{figure}

Now we have to integrate the next-to-leading order power spectra in
(\ref{P22rint}) and (\ref{P13rint}). As the linear power spectrum is
scale invariant, i.e. $\calP_\calR^{(11)} \propto k^0$, we have
$\calP_\calR^{(22)} \propto k^4$, $\calP_\calR^{(13a)} \propto k^4$
and $\calP_\calR^{(13b)} \propto k^2$.
Since we have worked in the large scale (super-horizon) limit, we
cannot integrate over the whole range of $q$ but
we have to introduce a cutoff in the maximum of $q$. As our basic
perturbation equations are valid only in the large scale limit, our
analysis is not valid near and inside the Hubble horizon scale $k_H
= a H$. Thus we may set $q_{\rm max} = k_H$, which gives
\begin{equation}\label{UVcutoff}
r_\text{max} = \frac{aH}{k} \, .
\end{equation}
Note that as this bound itself is $k$ dependent, the bound introduces
additional scale dependence. For infrared side we just take a
conservative range
\begin{equation}
r_\text{min} = 10^{-1000} \frac{aH}{k} \, .
\end{equation}
Although we also have logarithmic divergences of the integrals in
$r\to0$ limit, the infrared cutoff does not affect the result
appreciably (but see the discussion below). The
resulting power spectra under such scale dependent bounds are shown
in Figure~\ref{fig:Ptotal}.
The result apparently shows that non-linear contributions are completely negligible compared with the linear contribution in the observationally relevant scales.
Note that similar conclusions hold for the power spectrum of the
field fluctuation in the uniform curvature gauge $\calP_{\delta\phi}$~\cite{Sloth:2006az}.

Our result shows that the effect of leading order non-linear terms to the primordial power spectrum due to a single minimally coupled scalar field is completely negligible under our assumptions of the large scale and the slow-roll. Part of the reason can be found in the $[k/(aH)]^2$ suppression factor for the nonlinear terms in (\ref{3th_order_eq_ls}) and (\ref{Fourier_eq}). The effect of $[k/(aH)]^2$ suppression factor can be either regarded as large scale suppression occurred during the exponential expansion, or rapid decaying in time during the same expansion. That is, during the accelerated expansion a comoving scale rapidly becomes super-horizon scale. Since our leading order non-linear terms already have $[k/(aH)]^2$ terms, the non-linear contributions are suppressed during the accelerated stage as the evolution drives the comoving scales to outside the horizon.

We make a brief discussion on the integrands of the non-linear corrections.
In the limit $r\to0$, the leading term of each integrand is $1/r$
and thus is logarithmically divergent. More specifically, if we restrict ourselves to a box of
size $L=1/(aH)$, we have terms with $\log \left( kL \right)$~\cite{logboxsize}.
Thus, if we push the size of the box to literally infinity, we face a
divergence coming from infrared region~\cite{Seery:2010kh}. Indeed,
we have checked that as we push the infrared cutoff towards larger
and larger scales, the contributions of the non-linear corrections
increase. This infrared divergence may be removed with appropriate
manipulations, e.g. boundary effects~\cite{IRresolution}.

Conversely, we can find that the integrals in $\calP_\calR^{(13)}$
diverge in the large $r$ limit. This should be regarded as the
breakdown of our approximation near and inside the horizon scale.
Thus the ultraviolet cutoff we have chosen in (\ref{UVcutoff}),
horizon scale cutoff, is the maximum value of $r$ we can have:
beyond this value, we are probing sub-horizon regime where our
approximation is invalid. We may choose more conservative cutoff,
for example $10^{-1}aH/k$. The net effect of pushing $r_\text{max}$
to a smaller value is to suppress the non-linear corrections, since as
mentioned above the integrands in $\calP_\calR^{(13)}$ become bigger
at larger $r$.

\section{Conclusions}
\label{sec:conclusions}

In this note, we have studied the non-linear corrections to the
power spectrum of the comoving curvature perturbation produced during single
field slow-roll inflation. All the scalar perturbations in the
metric and the inflaton field are described in terms of the gauge
invariant comoving curvature perturbation $\calR$. If $\calR$ to the
linear order is Gaussian, we need up to third order perturbation to
describe the leading non-linear contributions to the power spectrum.
Under the assumptions of large scale limit and slow-roll approximation,
we have solved the equation of $\calR$ perturbatively
up to third order. Using these solutions, we have computed the
power spectrum $\calP_\calR$ including the leading non-linear
corrections. The resulting power spectrum is, on super-horizon
scales, dominated by the linear contribution $\calP_\calR^{(11)}$,
and the non-linear corrections are negligible\footnote{The
non-linear perturbations in the gradient expansion~\cite{NLdeltaN}
should be very closely related to our calculations in this note. We
would like to address this point in a separate report.}. Our study
indicates that we can safely rely on linear cosmological
perturbation theory to study power spectrum originated from quantum
fluctuations.

\subsection*{Acknowledgement}

We thank Donghui Jeong, Misao Sasaki, Martin Sloth and Takahiro Tanaka for useful conversations.
J.G.\ is grateful to the Yukawa Institute for
Theoretical Physics, Kyungpook National University
and Korea Astronomy and Space Science Institute for hospitality
where part of this work was carried out.
J.G.\ was supported in part
by a VIDI and a VICI Innovative Research Incentive Grant from the
Netherlands Organisation for Scientific Research (NWO) and a Korean-CERN fellowship.
H.N.\ was
supported by Mid-career Research Program through National Research
Foundation funded by the MEST (No.\ 2010-0000302).
J.H.\ was
supported by the Korea Research Foundation Grant funded by the
Korean Government (KRF-2008-341-C00022).

\end{document}